# Selection of Identifiability Criteria for Total Effects by using Path Diagrams

**Manabu Kuroki**
Department of Systems Innovation
Graduate School of Engineering Science
Osaka University
Machikaneyama-cho, Toyonaka, Osaka
mkuroki@sigmath.es.osaka-u.ac.jp

**Zhihong Cai**
Department of Biostatistics
Graduate School of Public Health
Kyoto University
Yoshida, Konoe-cho, Sakyo-ku, Kyoto
cai@pbh.med.kyoto-u.ac.jp

**Abstract**

Pearl has provided the back door criterion, the front door criterion and the conditional instrumental variable (IV) method as identifiability criteria for total effects. In some situations, these three criteria can be applied to identifying total effects simultaneously. For the purpose of increasing estimating accuracy, this paper compares the three ways of identifying total effects in terms of the asymptotic variance, and concludes that in some situations the superior of them can be recognized directly from the graph structure.

## 1 INTRODUCTION

Statistical causal analysis using directed graphs, investigated by many researchers (e.g. Wright, 1934; Bollen, 1989), is a beneficial tool for combining qualitative causal knowledge with statistical data. Pearl (2000) has in recent years developed a new framework of statistical causal analysis from these past causal studies, which enables us to clarify the meaning of "causality" in practical sciences. The central aim of statistical causal analysis is to evaluate a causal effect of a treatment variable $X$ on a response variable $Y$ through both qualitative causal knowledge and statistical data.

The total effect is a measure for evaluating causal effects, and can be interpreted as the change of the mean of the $Y$ when the $X$ is changed by one unit through an external intervention. In order to identify total effects, Pearl and his colleagues have provided the back door criterion and the front door criterion as identifiability criteria for total effects, when our causal knowledge is encoded in the form of a directed acyclic graph and the corresponding linear structural equation model (e.g. Pearl, 2000). Here, "identifiable" means that a total effect can be determined uniquely from the covariances of observed variables. Recently, they have developed the conditional instrumental variable (IV) method for identifying total effects (Brito and Pearl, 2002) as the generalization of the IV method (Bowden and Turkington, 1984).

Kuroki (2000) investigated the influence of selecting different sets of mediating variables which satisfy the front door criterion. Kuroki and Miyakawa (2003) and Miyakawa and Kuroki (1999) discussed the difference of the asymptotic variances of total effects when selecting various sets of covariates that are satisfied by the back door criterion, and showed that such difference can be recognized from the graph structure. In addition, Kuroki et al. (2003b) showed that in some cases a set of covariates used for the conditional IV method also satisfies the back door criterion. In such circumstances, the choice of the criterion to identify total effects becomes an essential problem. Although a great deal of effort has been devoted to establishing identifiability criteria for total effects, there has been much less discussion of the problem of which estimator to select when several are available. Since the estimating accuracy of total effects depends on the identifiability criterion, we discuss the selection problem of "the identifiability criteria for total effects" in terms of the asymptotic variance of total effects.

In this paper, we consider a case in which cause-effect relationships among variables are described by a directed acyclic graph and the corresponding linear structural equation model. We review the difference of the asymptotic variances of total effects when selecting various sets of covariates that are satisfied by the conditional IV method. Then, we compare the back door criterion and the conditional IV method by the asymptotic variances of their total effects, and obtain the result that the superior criterion can be identified solely from graph structure, without evaluating the precision of the total effects. Finally, numerical experiment and applicable example are provided to il-lustrate our results.

## 2 IDENTIFIABILITY CRITERIA FOR TOTAL EFFECTS

### 2.1 PATH DIAGRAM

In causal analysis, a directed acyclic graph that represents cause-effect relationships is called a path diagram. A directed graph is a pair $G = (\boldsymbol{V}, \boldsymbol{E})$, where $\boldsymbol{V}$ is a finite set of vertices and the set $\boldsymbol{E}$ of arrows is a subset of the set $\boldsymbol{V} \times \boldsymbol{V}$ of ordered pairs of distinct vertices. Graph theoretic terminology used in this paper is summarized in the Appendix.

**DEFINITION 1**

Suppose a directed acyclic graph $G = (\boldsymbol{V}, \boldsymbol{E})$ with set $\boldsymbol{V} = \{V_1, V_2, \cdots, V_n\}$ of variables is given. The graph $G$ is called a path diagram, when each child-parent family in the graph $G$ represents a linear structural equation model

$$V_i = \sum_{V_j \in \mathrm{pa}(V_i)} \alpha_{v_i v_j} V_j + \epsilon_{v_i} \qquad i = 1, \ldots, n, \quad (1)$$

where $\mathrm{pa}(V_i)$ denotes a set of parents of $V_i$ in $G$ and $\epsilon_{v_1}, \ldots, \epsilon_{v_n}$ are assumed to be independent and normally distributed with mean 0. In addition, $\alpha_{v_i v_j}(\neq 0)$ is called a path coefficient. □

The conditional independence induced from a set of equations (1) can be obtained from the graph $G$ according to the d-separation (Pearl, 2000).

**DEFINITION 2**

Let $\{X, Y\}$ and $\boldsymbol{Z}$ be disjoint sets of variables representing the vertices in a directed acyclic graph $G$. $\boldsymbol{Z}$ is said to d-separate $X$ from $Y$ in the graph $G$, if $\boldsymbol{Z}$ blocks every path between $X$ and $Y$. A path $p$ is said to be blocked by a (possibly empty) set $\boldsymbol{Z}$ if either of the following holds:

(1) $p$ contains at least one non-collider that is in $\boldsymbol{Z}$.

(2) $p$ contains at least one collider that is not in $\boldsymbol{Z}$ and has no descendant in $\boldsymbol{Z}$. □

When $\boldsymbol{Z}$ d-separates $X$ from $Y$ in a path diagram $G$, $X$ is conditionally independent of $Y$ given $\boldsymbol{Z}$ in the corresponding linear structural equation model (e.g. Spirtes et al., 1993).

Let $\sigma_{xy \cdot z} = cov(X, Y|\boldsymbol{Z})$, $\sigma_{yy \cdot z} = var(Y|\boldsymbol{Z})$, $\rho_{xy \cdot z} = cor(X, Y|\boldsymbol{Z}) = \sigma_{xy \cdot z}/\sqrt{\sigma_{xx \cdot z}\sigma_{yy \cdot z}}$ and $\beta_{yx \cdot z} = \sigma_{xy \cdot z}/\sigma_{xx \cdot z}$. Let $\boldsymbol{\beta}_{yz \cdot x}$ and $\boldsymbol{\sigma}_{xz}$ be vectored versions of $\beta_{yz \cdot x}$ and $\sigma_{xz}$, respectively. Further, let $\boldsymbol{\Sigma}_{zz \cdot x}$ be the conditional covariance matrix of $\boldsymbol{Z}$ given $X$. When $\boldsymbol{Z}$ is an empty set, $\boldsymbol{Z}$ is omitted from these arguments. The similar notations are used for other parameters.

When $\boldsymbol{Z}$ d-separates $X$ from $Y$ in a path diagram $G$, then $\sigma_{xy \cdot z} = \beta_{yx \cdot z} = 0$ and $\boldsymbol{\beta}_{yz \cdot x} = \boldsymbol{\beta}_{yz}$ (e.g. Spirtes, Glymour and Schienes, 1993).

### 2.2 IDENTIFIABILITY CRITERIA

In this section, we introduce the back door criterion (e.g. Pearl, 1995) and the conditional instrumental variable (IV) method (Brito and Pearl, 2002) as the identifiability criteria for total effects. Here, a total effect $\tau_{yx}$ of $X$ on $Y$ is defined as the total sum of the products of the path coefficients on the sequence of arrows along all directed paths from $X$ to $Y$. When a total effect can be determined uniquely from the correlation parameters of observed variables, it is said to be identifiable, that is, it can be estimated consistently. Let $G_{\underline{X}}$ be the graph obtained by deleting from a graph $G$ all arrows emerging from vertices in $\boldsymbol{X}$, and $G_{\overline{X}}$ be the graph obtained by deleting from a graph $G$ all arrows pointing to vertices in $\boldsymbol{X}$.

**DEFINITION 3**

Let $\{X, Y\}$ and $\boldsymbol{S}$ be disjoint subsets of $\boldsymbol{V}$ in a directed acyclic graph $G$. If a set $\boldsymbol{S}$ of variables satisfies the following conditions relative to an ordered pair $(X, Y)$ of variables, then $\boldsymbol{S}$ is said to satisfy the back door criterion relative to $(X, Y)$.

1. No vertex in $\boldsymbol{S}$ is a descendant of $X$, and

2. $\boldsymbol{S}$ d-separates $X$ from $Y$ in $G_{\underline{X}}$. □

If a set $\boldsymbol{S}$ of observed variables satisfies the back door criterion relative to $(X, Y)$ in a path diagram $G$, then the total effect $\tau_{yx}$ of $X$ on $Y$ is identifiable, and is given by the formula $\beta_{yx \cdot s}$ (Pearl, 2000). When $\beta_{yx \cdot s}$ is estimated by the ordinary least square estimator $\hat{\beta}_{yx \cdot s}$, the asymptotic variance of $\hat{\beta}_{yx \cdot s}$ is given by

$$a.var(\hat{\beta}_{yx \cdot s}) = \frac{1}{n}\frac{\sigma_{yy \cdot xs}}{\sigma_{xx \cdot s}},$$

where $a.var(\cdot)$ indicates the asymptotic variance and $n$ is the sample size.

**DEFINITION 4**

If a set $\boldsymbol{T} \cup \{Z\}$ of variables satisfies the following conditions relative to an ordered pair $(X, Y)$ of variables in a directed acyclic graph $G$, then $Z$ is said to be a conditional instrumental variable (IV) given $\boldsymbol{T}$ relative to $(X, Y)$.

1. A set $\boldsymbol{T}$ of variables is a subset of nondescendants of $X$ or $Y$ in $G$, and

2. $\boldsymbol{T}$ d-separates $Z$ from $Y$ but not from $X$ in $G_{\underline{X}}$. □

If an observed variable $Z$ is a conditional IV given a set $T$ of observed variables relative to $(X, Y)$, then the total effect $\tau_{yx}$ of $X$ on $Y$ is identifiable, and is given by $\sigma_{yz \cdot t}/\sigma_{xz \cdot t}$ (Brito and Pearl, 2002). When $T$ is an empty set in Definition 4, $Z$ is called an instrumental variable (IV) (Bowden and Turkington, 1984).

When the sample conditional covariances $\hat{\sigma}_{yz \cdot t}$ and $\hat{\sigma}_{xz \cdot t}$ of the conditional covariances $\sigma_{yz \cdot t}$ and $\sigma_{xz \cdot t}$ are used to estimate the total effect $\tau_{yx}$, by the delta method (Anderson, 1986), the asymptotic variance of $\hat{\sigma}_{yz \cdot t}/\hat{\sigma}_{xz \cdot t}$ is given by

$$a.var\left(\frac{\hat{\sigma}_{yz \cdot t}}{\hat{\sigma}_{xz \cdot t}}\right) = \frac{1}{n}\frac{\sigma_{yy \cdot t}/\sigma_{xx \cdot t} - 2\beta_{yx \cdot t}\tau_{yx} + \tau_{yx}^2}{\rho_{xz \cdot t}^2} \quad (2)$$

(Kuroki et al, 2003b).

## 3 SELECTION OF IDENTIFIABILITY CRITERIA

### 3.1 LEMMA

To clarify the difference between the identifiability criteria for total effects in terms of the asymptotic variance of total effects, we introduce the following lemmas:

**LEMMA 1**

When $\{X, Y\} \cup S \cup T$ are normally distributed,

$$\beta_{yx \cdot s} = \beta_{yx \cdot st} + \beta_{yt \cdot xs}\beta_{tx \cdot s}, \quad (3)$$

and

$$\begin{aligned}
\sigma_{yy \cdot xs} &= \sigma_{yy} - \beta_{yx \cdot s}^2 \sigma_{xx} - 2\beta_{yx \cdot s}\beta_{ys \cdot x}\sigma_{sx} \\
&\quad - \beta_{ys \cdot x}\Sigma_{ss}\beta'_{ys \cdot x}, \quad (4) \\
&= \sigma_{yy \cdot x} - \beta_{ys \cdot x}\Sigma_{ss \cdot x}\beta'_{ys \cdot x}. \quad (5)
\end{aligned}$$

$\square$

**PROOF**

Equations (3) and (4) are the results of Cochran (1938) and Kuroki and Miyakawa (2003), respectively. Based on equations (3) and (4), we can obtain a new result

$$\begin{aligned}
\sigma_{yy \cdot xs} &= \sigma_{yy} - \sigma_{xx}(\beta_{yx \cdot s} + \beta_{ys \cdot x}\beta_{sx})^2 \\
&\quad + \sigma_{xx}(\beta_{ys \cdot x}\beta_{sx})^2 - \beta_{ys \cdot x}\Sigma_{ss}\beta'_{ys \cdot x} \\
&= \sigma_{yy} - \sigma_{xx}\beta_{yx}^2 \\
&\quad - \beta_{ys \cdot x}(\Sigma_{ss} - \sigma_{xx}\beta_{sx}\beta'_{sx})\beta'_{ys \cdot x} \\
&= \sigma_{yy \cdot x} - \beta_{ys \cdot x}\Sigma_{ss \cdot x}\beta'_{ys \cdot x}.
\end{aligned}$$

Q.E.D.

The following lemma was given by Wermuth (1989):

**LEMMA 2**

When $\{X, Y\} \cup S \cup T$ are normally distributed, if $T$ is conditionally independent of $X$ given $S$ or $Y$ is conditionally independent of $T$ given $\{X\} \cup S$, then $\beta_{yx \cdot st} = \beta_{yx \cdot s}$ holds true. In addition, if $T$ is conditionally independent of $X$ given $S$, then $\sigma_{yy \cdot xst} = \sigma_{yy \cdot xs}$ holds true. $\square$

The following lemma is a slight generalization of Miyakawa and Kuroki (1999):

**LEMMA 3**

Suppose that $T$ d-separates $X$ from $S$, and $S \cup \{X\}$ d-separates $T$ from $Y$ in a path diagram $G$. When $\{X, Y\} \cup S \cup T$ are normally distributed,

$$\frac{\sigma_{yy \cdot xs}}{\sigma_{xx \cdot s}} \leq \frac{\sigma_{yy \cdot xst}}{\sigma_{xx \cdot st}} \leq \frac{\sigma_{yy \cdot xt}}{\sigma_{xx \cdot t}}.$$

$\square$

**PROOF**

From Lemmas 1 and 2, we can obtain $\beta_{xs} = \beta_{xs \cdot t} + \beta_{xt \cdot s}\beta_{ts} = \beta_{xt}\beta_{ts}$ and $\beta_{yx \cdot t} = \beta_{yx \cdot s}$, where $\beta_{ts}$ is the regression coefficient matrix whose $(i, j)$ element is the regression coefficient of $s_j$ in the regression model of $T_i \in T' = (T_1, \cdots, T_m)$ on $s = (s_1, \cdots, s_p)$.

Then, by Lemma 1,

$$\sigma_{xx \cdot s} - \sigma_{xx \cdot t} = \beta_{xt}\Sigma_{tt}\beta'_{xt} - \beta_{xs}\Sigma_{ss}\beta'_{xs} = \beta_{xt}\Sigma_{tt \cdot s}\beta'_{xt} \geq 0.$$

In addition, since $\beta_{yt \cdot x} = \beta_{yt \cdot xs} + \beta_{ys \cdot xt}\beta_{st \cdot x} = \beta_{ys \cdot x}\beta_{st \cdot x}$, $\sigma_{yy \cdot xt} = \sigma_{yy \cdot x} - \beta_{yt \cdot x}\Sigma_{tt \cdot x}\beta'_{yt \cdot x}$ and $\sigma_{yy \cdot xs} = \sigma_{yy \cdot x} - \beta_{ys \cdot x}\Sigma_{ss \cdot x}\beta'_{ys \cdot x}$, according to Lemma 1,

$$\begin{aligned}
\sigma_{yy \cdot xt} - \sigma_{yy \cdot xs} &= \beta_{ys \cdot x}\Sigma_{ss \cdot x}\beta'_{ys \cdot x} - \beta_{yt \cdot x}\Sigma_{tt \cdot x}\beta'_{yt \cdot x} \\
&= \beta_{ys \cdot x}\Sigma_{ss \cdot xt}\beta'_{ys \cdot x} \geq 0.
\end{aligned}$$

Further, $\sigma_{yy \cdot xst} = \sigma_{yy \cdot xs}$ and $\sigma_{xx \cdot st} = \sigma_{xx \cdot t}$ according to Lemma 2. Thus, the lemma follows from these results directly. Q.E.D.

Consider the path diagram shown in Fig.1 as an example. Since $T$ d-separates $X$ from $S$, $X$ is conditionally independent of $S$ given $T$. Similarly, since $S \cup \{X\}$ d-separates $T$ from $Y$, $Y$ is conditionally independent of $T$ given $S \cup \{X\}$. Since both $T$ and $S$ satisfy the back door criterion relative to $(X, Y)$ and the condition described in Lemma 3, both $\hat{\beta}_{yx \cdot t}$ and $\hat{\beta}_{yx \cdot s}$ are estimates of the total effect $\tau_{yx}$. Regarding the asymptotic variance of the total effect $\tau_{yx}$ of $X$ on $Y$, we can obtain

$$a.var(\hat{\beta}_{yx \cdot s}) = \frac{\sigma_{yy \cdot xs}}{n\sigma_{xx \cdot s}} \leq a.var(\hat{\beta}_{yx \cdot t}) = \frac{\sigma_{yy \cdot xt}}{n\sigma_{xx \cdot t}}.$$

That is, the asymptotic variance when $S$ is used is smaller than that when $T$ is used under the conditions above.

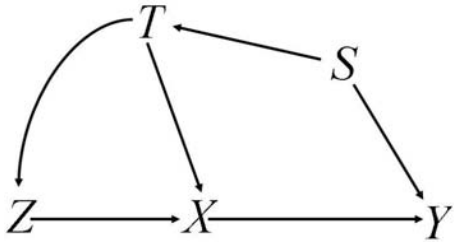

Fig. 1: Path Diagram (1)

## 3.2 CONDITIONAL IV SELECTION:REVIEW

Consider the path diagram shown in Fig.2. Both $Z_1$

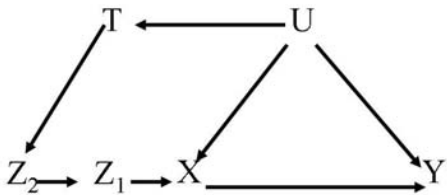

Fig. 2: Path Diagram (2)

and $Z_2$ are the conditional IVs given $\boldsymbol{T}$ relative to $(X,Y)$, and $\{Z_1\}\cup\boldsymbol{T}$ d-separates $X$ from $Z_2$. Hence, there is no unique conditional IV. Regarding this problem, Kuroki et al. (2003b) stated the following proposition:

### PROPOSITION 1

Suppose that both $Z_1$ and $Z_2$ are conditional IVs given $\boldsymbol{T}$ relative to $(X,Y)$. If $\{Z_1\}\cup\boldsymbol{T}$ d-separates $X$ from $Z_2$ in a path diagram $G$, and $\{X,Y,Z_1,Z_2\}\cup\boldsymbol{T}$ are normally distributed,

$$a.var(\frac{\hat{\sigma}_{yz_1 \cdot t}}{\hat{\sigma}_{xz_1 \cdot t}}) \leq a.var(\frac{\hat{\sigma}_{yz_2 \cdot t}}{\hat{\sigma}_{xz_2 \cdot t}}). \qquad (6)$$

□

**PROOF**
The proof of this proposition is presented in Kuroki et al. (2003b). Here, we outline it as follows: $\rho^2_{xz \cdot t} = \sigma^2_{xz \cdot t}/(\sigma_{xx \cdot t}\sigma_{zz \cdot t}) = \beta^2_{xz \cdot t}\sigma_{zz \cdot t}/\sigma_{xx \cdot t}$ holds true. Since $\{Z_1\}\cup\boldsymbol{T}$ d-separates $Z_2$ from $X$, the following equation can be obtained by applying the results of Cochran (1938) and Wermuth (1989):

$$\beta^2_{xz_2 \cdot t}\sigma_{z_2 z_2 \cdot t} = \beta^2_{xz_1 \cdot t}\beta^2_{z_1 z_2 \cdot t}\sigma_{z_2 z_2 \cdot t}$$

$$= \beta^2_{xz_1 \cdot t}\sigma_{z_1 z_1 \cdot t}\rho^2_{z_1 z_2 \cdot t} \leq \beta^2_{xz_1 \cdot t}\sigma_{z_1 z_1 \cdot t}.$$

Proposition 1 can be obtained by noting these results.
Q.E.D.

Regarding Fig.2, since $\{Z_1\}\cup\boldsymbol{T}$ d-separates $X$ from $Z_2$, equation (6) holds.

### 3.3 BACK DOOR CRITERION vs. CONDITIONAL IV METHOD:NEW RESULTS

In Fig.1, $\boldsymbol{S}$ satisfies the back door criterion relative to $(X,Y)$ and $Z$ is a conditional IV given $\boldsymbol{T}$ relative to $(X,Y)$. Since $\sigma_{yy \cdot xt} = \sigma_{yy \cdot t} - \beta^2_{yx \cdot t}\sigma_{xx \cdot t}$, we have $\sigma_{yy \cdot xt}/\sigma_{xx \cdot t} = \sigma_{yy \cdot t}/\sigma_{xx \cdot t} - \beta^2_{yx \cdot t}$. Therefore, from Lemma 3,

$$a.var(\hat{\beta}_{yx \cdot s}) - a.var\left(\frac{\hat{\sigma}_{yz \cdot t}}{\hat{\sigma}_{xz \cdot t}}\right)$$
$$= \frac{1}{n}\left(\frac{\sigma_{yy \cdot xs}}{\sigma_{xx \cdot s}} - \frac{\sigma_{yy \cdot xt}/\sigma_{xx \cdot t} + (\beta_{yx \cdot t} - \tau_{yx})^2}{\rho^2_{xz \cdot t}}\right) \leq 0,$$

since $\boldsymbol{T}$ d-separates $X$ from $\boldsymbol{S}$, and $\boldsymbol{S}\cup\{X\}$ d-separates $\boldsymbol{T}$ from $Y$. Therefore, the following proposition holds true:

### PROPOSITION 2

Suppose that $Z$ is a conditional IV given $\boldsymbol{T}$ relative to $(X,Y)$ and that $\boldsymbol{S}$ satisfies the back door criterion relative to $(X,Y)$ in a path diagram $G$. When $\{X,Y\}\cup\boldsymbol{S}\cup\boldsymbol{T}$ are normally distributed, if $\boldsymbol{T}$ d-separates $X$ from $\boldsymbol{S}$, and $\boldsymbol{S}\cup\{X\}$ d-separates $\boldsymbol{T}$ from $Y$ in the $G$, then

$$a.var(\hat{\beta}_{yx \cdot s}) \leq a.var\left(\frac{\hat{\sigma}_{yz \cdot t}}{\hat{\sigma}_{xz \cdot t}}\right).$$

□

Proposition 2 also holds when $\boldsymbol{T}$ is identical to $\boldsymbol{S}$. Proposition 2 shows that if a set of covariates used for the conditional IV method satisfies the back door criterion, then the asymptotic variance of a total effect when using the back door criterion is smaller than its conditional IV method counterpart, which is not dependent on the selection of the conditional IVs.

When the assumptions stated in Proposition 2 are not satisfied, the asymptotic variances of the total effect based on the back door criterion and on the conditional IV method cannot be compared solely from the graph structure. As an example in which the assumptions stated in Proposition 2 are not satisfied, consider the following covariance matrix corresponding to Fig.1.

$$\begin{pmatrix} \sigma_{yy} & \sigma_{yx} & \sigma_{yz} & \sigma_{yt} & \sigma_{ys} \\ \sigma_{xy} & \sigma_{xx} & \sigma_{xz} & \sigma_{xt} & \sigma_{xs} \\ \sigma_{zy} & \sigma_{zx} & \sigma_{zz} & \sigma_{zt} & \sigma_{zs} \\ \sigma_{ty} & \sigma_{tx} & \sigma_{tz} & \sigma_{tt} & \sigma_{ts} \\ \sigma_{sy} & \sigma_{sx} & \sigma_{sz} & \sigma_{st} & \sigma_{ss} \end{pmatrix}$$

Table 1: Simulation Results

|  | Variable | $n=20$ | $n=40$ | $n=60$ | $n=80$ | $n=100$ |
|---|---|---|---|---|---|---|
| Back door criterion | $T$ | 0.097 | 0.043 | 0.028 | 0.020 | 0.016 |
|  | $S$ | 0.036 | 0.016 | 0.010 | 0.008 | 0.006 |
| Conditional IV | $T$ | 0.113 | 0.057 | 0.034 | 0.025 | 0.020 |
|  |  | (0.081) | (0.041) | (0.027) | (0.020) | (0.016) |
|  | $S$ | 0.062 | 0.027 | 0.017 | 0.011 | 0.009 |
|  |  | (0.045) | (0.022) | (0.015) | (0.011) | (0.009) |

$$= \begin{pmatrix} 1.000 & 0.277 & 0.184 & 0.248 & 0.626 \\ 0.277 & 1.000 & 0.800 & 0.640 & 0.128 \\ 0.184 & 0.800 & 1.000 & 0.200 & 0.040 \\ 0.248 & 0.640 & 0.200 & 1.000 & 0.200 \\ 0.626 & 0.128 & 0.040 & 0.200 & 1.000 \end{pmatrix} \quad (7)$$

Suppose that the pair $(S, Z)$ is used for the conditional IV method and $T$ is used for the back door criterion, which is a situation where the assumptions stated in Proposition 2 do not hold true. Based on this covariance matrix, $n \times a.var(\hat{\beta}_{yx\cdot t}) = 1.550 > n \times a.var(\hat{\sigma}_{yz\cdot s}/\hat{\sigma}_{xz\cdot s}) = 0.900$. That is, when $T$ is used for the back door criterion, the asymptotic variance of total effects is larger than that when the pair $(S, Z)$ is used for the conditional IV method. This example shows that the back door criterion does not always make the asymptotic variance smaller than the conditional IV method.

On the other hand, when the pair $(T, Z)$ is used for the conditional IV method and $S$ is used for the back door criterion, since that situation satisfies the assumptions stated in Proposition 2, $a.var(\hat{\beta}_{yx\cdot s}) \le a.var(\hat{\sigma}_{yz\cdot t}/\hat{\sigma}_{xz\cdot t})$ can be recognized from the graph structure without evaluating the precision of total effects.

## 4 FRONT DOOR CRITERION

The front door criterion is also a graphical identifiability criterion for total effects, which is described as follows (Pearl, 1995, 2000):

**DEFINITION 5**

Let $\{X, Y\}, \boldsymbol{Z}$ be disjoint subsets of $\boldsymbol{V}$ in a directed acyclic graph $G$. If a set $\boldsymbol{Z}$ of variables satisfies the following conditions relative to an ordered pair $(X, Y)$ of variables, then $\boldsymbol{Z}$ is said to satisfy the front door criterion relative to $(X, Y)$.

1. $\boldsymbol{Z}$ d-separates $X$ from $Y$ in $G_{\bar{X}}$.

2. an empty set d-separates $X$ from any vertex in $\boldsymbol{Z}$ in $G_{\underline{X}}$.

3. $X$ d-separates any vertex in $\boldsymbol{Z}$ from $Y$ in $G_{\underline{Z}}$. □

If a set $\boldsymbol{Z}' = (Z_1, \cdots, Z_r)$ of observed variables satisfies the front door criterion relative to $(X, Y)$ in a path diagram $G$, then the total effect $\tau_{yx}$ of $X$ on $Y$ is identifiable, and is given by the formula $\beta_{yz\cdot x}\beta_{zx}$ (Pearl, 2000). When the ordinary least square estimators $\hat{\beta}_{yz\cdot x}$ and $\hat{\beta}_{zx}$ of the regression coefficient $\beta_{yz\cdot x}$ and $\beta_{zx}$ are used to estimate the total effect $\tau_{yx}$, Kuroki (2000) provided the exact variance

$$\begin{aligned} &var(\hat{\beta}_{yz\cdot x}\hat{\beta}_{zx}) \\ &= \left( \frac{1}{(n-r-3)\sigma_{xx\cdot z}} - \frac{1}{(n-3)\sigma_{xx}} \right) \sigma_{yy\cdot xz} \\ &\quad + \frac{1}{(n-3)\sigma_{xx}} \beta_{yz\cdot x} \Sigma_{zz\cdot x} \beta'_{yz\cdot x}. \end{aligned}$$

In case where there are some mediating variables satisfying the front door criterion, Kuroki (2000) discussed the difference between the mediating variables on the basis of the asymptotic variance, and indicated that it is difficult to recognize the difference from graph structure. In general, the difference between the front door criterion and the other two criteria on the basis of the asymptotic variance can not be reconciled from the graph structure.

## 5 NUMERICAL EXPERIMENT

In this section, we will examine the variance of the total effect in sample sizes $n = 20 - 100$. The path diagram and the correlation matrix used in the study are given in Fig.1 and equation (7), respectively.

We simulated $n$ samples from a multivariate normal distribution with the correlation matrix (7). Then, we evaluated the empirical variances of the total effects 1000 times using $n = 20 - 100$. Table 1 reports the variances of the total effects when the back door criterion and the conditional IV method are selected. When $\boldsymbol{T}' = (T_1, \cdots, T_q)$ satisfies the back door criterion relative to $(X, Y)$, the variance of the total effect $\tau_{yx}$ is given by

$$var(\hat{\beta}_{yx\cdot t}) = \frac{\sigma_{yy\cdot xt}}{(n-q-3)\sigma_{xx\cdot t}}$$

(Miyakawa and Kuroki, 1999).

The values in the parentheses represent the asymptotic variance of the total effect, which can be calculated by equation (2). From this table, we draw the following conclusions:

1) Comparing the asymptotic variances with the variances regarding the conditional IV methods, the ratios $var(\hat{\sigma}_{yz \cdot t}/\hat{\sigma}_{xz \cdot t})/\ a.var(\hat{\sigma}_{yz \cdot t}/\hat{\sigma}_{xz \cdot t})$ and $var(\hat{\sigma}_{yz \cdot s}/\hat{\sigma}_{xz \cdot s})/\ a.var(\hat{\sigma}_{yz \cdot s}/\hat{\sigma}_{xz \cdot s})$ are larger than 1.4 when $n = 20$ and close to 1 when $n = 100$ in each case.

2) Regarding the same covariate, the variance when the back door criterion is selected is smaller than the variance when the conditional IV method is selected in all $n$, which are consistent with the results of Proposition 2.

3) The variance when $T$ is used for the back door criterion is larger than the variance when $S$ is used for the conditional IV method in all $n$, which is consistent with the description in section 3.3.

4) The variance ratios $var(\hat{\beta}_{yx \cdot t})/var(\hat{\beta}_{yx \cdot s})$ are larger than 2.6 when the back door criterion is applied, but $var(\hat{\sigma}_{yz \cdot t}/\hat{\sigma}_{xz \cdot t})/var(\hat{\sigma}_{yz \cdot s}/\hat{\sigma}_{xz \cdot s})$ are smaller than 2.2 when the conditional IV method is applied for all $n$. Therefore, there is a noticeable difference between the selected covariates as well as between the selected criteria.

## 6 APPLICATION

The above results are applicable to analyze the data obtained from a study on setting up painting conditions of car bodies, reported by Okuno et al. (1986). The data was collected with the purpose of setting up the process conditions, in order to increase transfer efficiency. The size of the sample is 38 and the variables of interest, each of which has zero mean and variance one, are the following:

Painting Condition : Dilution Ratio ($X_1$),
   Degree of Viscosity ($X_2$), Painting Temperature ($X_8$)

Spraying Condition : Gun Speed ($X_3$),
   Spray Distance ($X_4$), Atomizing Air Pressure ($X_5$),
   Pattern Width ($X_6$), Fluid Output ($X_7$)

Environment Condition : Temperature ($X_9$),
   Degree of Moisture ($X_{10}$)

Response: Transfer Efficiency ($Y$)

Concerning this process, Kuroki et al. (2003a) presented the path diagram shown in Fig.3 (for the detail, see Kuroki et al., 2003a). Based on the path diagram, Kuroki et al. (2003a) presented the estimated correlation matrix shown in Table 2.

Here, $X_1, \cdots, X_6$ are considered to be controllable variables according to Okuno et al. (1986). It is supposed that $X_2$, $X_4$ and $X_6$ are taken as treatment variables from controllable variables in order to evaluate their total effects from nonexperimental data.

Table 3 shows the selected variables for estimating total effects. Treatment variables $X_2$, $X_4$ and $X_6$ are listed in the first column. The second column shows the conditional IVs corresponding to these treatment variables. The third and the fourth columns shows the set of covariates used in the conditional IV method and the set of covariates satisfying the back door criterion, respectively.

Regarding the effect of $X_2$, from Fig.3, it is obvious that the covariate $\{X_{10}\}$ used in the conditional IV method does not satisfy the back door criterion, but $\{X_8, X_{10}\}$ satisfies the back door criterion. Hence, the assumptions stated in Proposition 2 does not hold true. Therefore, it is difficult to judge the difference between $n \times a.var(\hat{\sigma}_{yx_1 \cdot x_{10}}/\hat{\sigma}_{x_1 x_2 \cdot x_{10}}) = 7.576$ and $n \times a.var(\hat{\beta}_{yx_2 \cdot x_8 x_{10}}) = 1.456$ from graph structure.

Regarding the effect of $X_4$, for any $X_i$ ($i = 1, 2, 3, 9, 10$), in case where a set $\{X_7, X_8\}$ of covariates is used for the conditional IV method and a covariate $\{X_9\}$ is used for the back door criterion, the differences between $n \times a.var(\hat{\sigma}_{yx_i \cdot x_7 x_8}/\hat{\sigma}_{x_4 x_i \cdot x_7 x_8})$ and $n \times a.var(\hat{\beta}_{yx_4 \cdot x_9}) = 0.810$ can not be recognized from graph structure, since the assumptions stated in Proposition 2 do not hold true. On the other hand, in case where a set $\{X_7, X_8\}$ of covariates used for the conditional IV method also satisfies the back door criterion relative to $(X_4, Y)$, from Proposition 2, $n \times a.var(\hat{\beta}_{yx_4 \cdot x_7 x_8}) = 0.709 \leq n \times a.var(\hat{\sigma}_{yx_i \cdot x_7 x_8}/\hat{\sigma}_{x_4 x_i \cdot x_7 x_8})$ can be recognized without the calculation of $X_i$. For example, we can obtain $n \times a.var(\hat{\sigma}_{yx_9 \cdot x_7 x_8}/\hat{\sigma}_{x_4 x_9 \cdot x_7 x_8}) = 5.273$ and $n \times a.var(\hat{\beta}_{yx_4 \cdot x_7 x_8}) = 0.615$ from Table 2, which is consistent with the result of Proposition 2.

In addition, since both $\{X_7, X_8\}$ and $\{X_9\}$ satisfy the back door criterion relative to $(X_4, Y)$, by letting $\boldsymbol{S} = \{X_7, X_8\}$ and $\boldsymbol{T} = \{X_9\}$, the assumptions stated in Lemma 3 hold true. Therefore, we can obtain $n \times a.var(\hat{\beta}_{yx_4 \cdot x_7 x_8}) = 0.615 \leq n \times a.var(\hat{\beta}_{yx_4 \cdot x_9}) = 0.810$ without the calculation of the asymptotic variance.

Regarding the effect of $X_6$, the covariate used for the conditional IV method also satisfies the back door criterion relative to $(X_6, Y)$ from Fig.3. Therefore, from Proposition 2, $n \times a.var(\hat{\beta}_{yx_6 \cdot x_4}) = 0.932 \leq n \times a.var(\hat{\sigma}_{yx_5 \cdot x_4}/\hat{\sigma}_{x_5 x_6 \cdot x_4}) = 3.963$ can be recognized without the calculation.

From Table 3, when the conditional IV method is used in order to estimate a total effect of $X_4$ on $Y$, there are several choices of conditional IVs given $\{X_7, X_8\}$. Since $\{X_7, X_8, X_9\}$ d-separates $X_4$ from $\{X_1, X_2, X_3, X_{10}\}$ in Fig.3, the asymptotic variance in case where $X_9$ is used as a conditional IV is smaller than that in case where other variables are used as conditional IVs. For example, we can obtain $n \times a.var(\hat{\sigma}_{yx_{10} \cdot x_7 x_8}/\hat{\sigma}_{x_4 x_{10} \cdot x_7 x_8}) = 26.3532$ and $n \times a.var(\hat{\sigma}_{yx_9 \cdot x_7 x_8}/\hat{\sigma}_{x_4 x_9 \cdot x_7 x_8}) = 5.2732$ from Table 2, which is consistent with the result of Proposition 1.

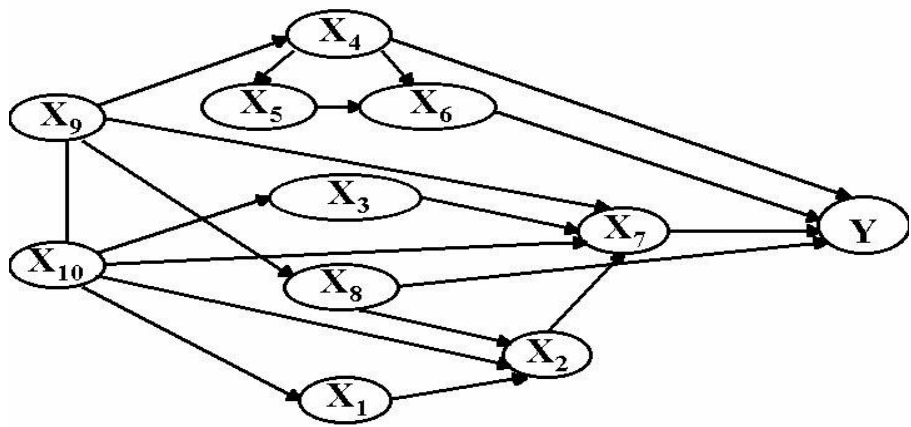

Fig.3 : Path Diagram (Kuroki et al., 2003a)

Table 2 : The estimated correlation matrix based on Fig.3 (Kuroki et al., 2003a)

|   | $X_1$ | $X_2$ | $X_3$ | $X_4$ | $X_5$ | $X_6$ | $X_7$ | $X_8$ | $X_9$ | $X_{10}$ | $Y$ |
|---|---|---|---|---|---|---|---|---|---|---|---|
| $X_1$ | 1.000 | -0.736 | -0.152 | 0.148 | 0.028 | -0.042 | 0.324 | 0.216 | 0.283 | -0.496 | -0.091 |
| $X_2$ | -0.736 | 1.000 | 0.210 | -0.331 | -0.063 | 0.095 | -0.479 | -0.684 | -0.635 | 0.684 | 0.326 |
| $X_3$ | -0.152 | 0.210 | 1.000 | -0.091 | -0.017 | 0.026 | 0.195 | -0.134 | -0.175 | 0.307 | 0.134 |
| $X_4$ | 0.148 | -0.331 | -0.091 | 1.000 | 0.191 | -0.286 | 0.184 | 0.397 | 0.521 | -0.298 | -0.614 |
| $X_5$ | 0.028 | -0.063 | -0.017 | 0.191 | 1.000 | 0.291 | 0.035 | 0.076 | 0.099 | -0.057 | -0.277 |
| $X_6$ | -0.042 | 0.095 | 0.026 | -0.286 | 0.291 | 1.000 | -0.053 | -0.114 | -0.149 | 0.085 | -0.250 |
| $X_7$ | 0.324 | -0.479 | 0.195 | 0.184 | 0.035 | -0.053 | 1.000 | 0.396 | 0.353 | -0.146 | -0.044 |
| $X_8$ | 0.216 | -0.684 | -0.134 | 0.397 | 0.076 | -0.114 | 0.396 | 1.000 | 0.761 | -0.435 | -0.493 |
| $X_9$ | 0.283 | -0.635 | -0.175 | 0.521 | 0.099 | -0.149 | 0.353 | 0.761 | 1.000 | -0.571 | -0.475 |
| $X_{10}$ | -0.496 | 0.684 | 0.307 | -0.298 | -0.057 | 0.085 | -0.146 | -0.435 | -0.571 | 1.000 | 0.283 |
| $Y$ | -0.091 | 0.326 | 0.134 | -0.614 | -0.277 | -0.250 | -0.044 | -0.493 | -0.475 | 0.283 | 1.000 |

Table 3 : The conditional IV method and the back door criterion

| Treatment | Conditional IV | Covariates for Conditional IV | Covariates for Back Door Criterion |
|---|---|---|---|
| $X_2$ | $\{X_1\}$ | $\{X_{10}\}$ | $\{X_8, X_{10}\}$ |
| $X_4$ | $\{X_1\},\{X_2\},\{X_3\}$ $\{X_9\},\{X_{10}\}$ | $\{X_7, X_8\}$ | $\{X_9\}, \{X_7, X_8\}$ |
| $X_6$ | $\{X_5\}$ | $\{X_4\}$ | $\{X_4\}$ |

Regarding other variables, it is difficult to recognize the difference between the conditional IVs from graph structure (Kuroki et al., 2003b).

## 7 CONCLUSION

This paper discussed the graphical selection problem for identifiability criteria of total effects for the purpose of increasing estimating accuracy. When there are several sets of covariates that satisfy the back door criterion, we can judge from graph structure the difference among various sets of covariates in terms of the asymptotic variance. This result is a generalization of Miyakawa and Kuroki (1999). When there are several conditional IVs, this paper reviewed the difference when selecting different conditional IVs. In addition, this paper clarified a graphical situation in which the back door criterion provides superior estimation accuracy of a total effect than the conditional IV method. The results of this paper help us judge from graph structure which criterion for total effects should be selected in order to increase estimating accuracy.

## 8 APPENDIX

A directed graph is a pair $G = (\boldsymbol{V}, \boldsymbol{E})$, where $\boldsymbol{V}$ is a finite set of vertices and the set $\boldsymbol{E}$ of arrows is a subset $\boldsymbol{V} \times \boldsymbol{V}$ of ordered pairs of distinct vertices. An arrow pointing from a vertex $a$ to a vertex $b$ indicates $(a, b) \in \boldsymbol{E}$ and $(b, a) \notin \boldsymbol{E}$. The arrow is said to emerge from $a$ and point to $b$. If there is an arrow pointing from $a$ to $b$, $a$ is said to be a parent of $b$, and $b$ a child of $a$.

A path between $a$ and $b$ is a sequence $a = a_0, a_1, \cdots, b = a_n$ of distinct vertices such as $(a_{i-1}, a_i) \in \boldsymbol{E}$ or $(a_i, a_{i-1}) \in \boldsymbol{E}$ for all $i = 1, 2, \cdots, n$. A directed path from $a$ to $b$ is a sequence $a = a_0, a_1, \cdots, b = a_n$ of distinct vertices such as $(a_{i-1}, a_i) \in \boldsymbol{E}$ and $(a_i, a_{i-1}) \notin \boldsymbol{E}$ for all $i = 1, \cdots, n$. If there exists a directed path from $a$ to $b$, $a$ is said to be an ancestor of $b$ and $b$ a descendant of $a$. When the set of descendants of $a$ is denoted as $de(a)$, the vertices in $\boldsymbol{V} \setminus (de(a) \cup \{a\})$ are said to be the nondescendants of $a$. If two arrows on a path point to $a$, then $a$ is said to be a collider; Otherwise, it is said to be a non-collider.

A directed path that begins and ends in the same vertex is said to be a cycle. If a directed graph has no cycles, then the graph is said to be a directed acyclic graph.

### ACKNOWLEGDEMENT

The comments of the reviewers on preliminary versions of this paper are acknowledged.